\documentclass[10pt]{article}

\usepackage{amsmath}
\usepackage{amssymb}

\usepackage{graphicx}

\usepackage{cite}

\usepackage{color} 

\makeatletter
\renewcommand{\@biblabel}[1]{\quad#1.}
\makeatother

\date{}

\begin{document}

\begin{flushleft}
{\Large
\textbf{Basic Experiment Planning via Information Metrics: the RoboMendel Problem}
}
\\
Christopher Lee$^{1,2,3,4}$,
Marc Harper$^{1}$
\\
\bf{1} Institute for Genomics and Proteomics, University of California, Los Angeles, CA, USA
\\
\bf{2} Dept. of Chemistry \& Biochemistry,  University of California, Los Angeles, CA, USA
\\
\bf{3} Dept. of Computer Science,  University of California, Los Angeles, CA, USA
\\
\bf{4} Molecular Biology Institute,  University of California, Los Angeles, CA, USA
\\
$\ast$ E-mail: Corresponding leec@chem.ucla.edu
\end{flushleft}

\begin{abstract}
In this paper we outline some mathematical questions that emerge
from trying to ``turn the scientific method into math''.
Specifically, we consider the problem of experiment planning
(choosing the best experiment to do next) in explicit
probabilistic and information theoretic terms.  We formulate
this as an information measurement problem; that is, we seek a
rigorous definition of an information metric to measure the likely information
yield of an experiment, such that maximizing the information
metric will indeed reliably choose the best experiment to
perform.  We present the surprising result that defining
the metric purely in terms of prediction power on \emph{observable}
variables yields a metric that can converge to the classical
mutual information $I(X;\Omega)$
measuring how informative the experimental
observation $X$ is about an underlying \emph{hidden} variable $\Omega$.
We show how the expectation potential information metric can
compute the ``information rate'' of an experiment as well its total
possible yield, and the information value of experimental controls.
To illustrate the utility of these
concepts for guiding fundamental scientific inquiry,
we present an extensive case study (RoboMendel) applying these metrics
to propose sequences of experiments for discovering the basic principles
of genetics.
\end{abstract}


\section{Introduction}
\label{exptplan:introduction}\label{exptplan:basic-experiment-planning-via-information-metrics-the-robomendel-problem}

\subsection{The RoboMendel Problem}
\label{exptplan:the-robomendel-problem}
We first outline an intuitive introduction to our example application problem
and its assumptions.  We consider the problem of a robot scientist
(whom we call ``RoboMendel'') assigned the task of improving his
theory of inheritance of traits, as measured by his ability to
predict the outcome of any possible cross between animals, plants and
other living creatures.  We will derive precise definitions (e.g. of
this ``prediction power'' metric) later; for the moment we focus
on an intuitive statement of the problem and its goal.  We assume
that RoboMendel begins roughly where Mendel began, namely:
\begin{itemize}
\item {} 
He starts with a simple ``like father, like son'' (LFLS) model of inheritance
that treats heritable variation as a set of distinct clusters
(i.e. species) in ``trait space''.  That is, each species has a
distinct distribution of trait values, and all members of that
species are drawn i.i.d. (independent and identically
distributed) from that multi-dimensional distribution.
Members of a
species can only give birth to new instances of the same species
(again drawn i.i.d. from its trait distribution).  The simplest
version of this model asserts that two parents (mother and father)
are required to give progeny, and that they must be from the same
species.

\item {} 
Beyond this, RoboMendel does not know any modern biology or technology,
and thus can only detect visible, macroscopic trait differences
that Mendel could see with his eyes.

\item {} 
Concretely, the set of all possible ``inheritance'' experiments
RoboMendel could do is simply the set of all possible crosses between
different creatures.  For example, he could try to cross
a mouse with a lion, or a pea plant with another pea plant, etc.

\item {} 
We define RoboMendel's goal as elucidating predictive laws of
inheritance, which we define as traits that in any individual
depend on the traits of its ancestors (as opposed to other factors
such as the weather).  For example, we can easily see that
the ``like father, like son'' model fits this definition, as
follows.  For each individual organism $i$, we associate
with it a trait vector $\vec X_i$ of its observed trait
values, and a hidden value $\Theta_i$ indicating what
species it belongs to.  Mathematically, the problem of
predicting an individual's traits ($p(\vec X_i|\Theta_i)$)
and the converse problem of inferring its species
($p(\Theta_i|\vec X_i)$) are
linked by Bayes' Law.  The main assertion of LFLS, namely
that $\Theta_{\text{child}}=\Theta_{\text{dad}}=\Theta_{\text{mom}}$,
indeed makes the parental traits
$\vec X_{\text{dad}},\vec X_{\text{mom}}$
the sufficient statistics for predicting
the child's traits $\vec X_{\text{child}}$,
so LFLS fits our definition of a ``predictive law of inheritance''.
We will later define precisely how this statement of a ``target problem''
is translated into an information metric.

\item {} 
RoboMendel is initially shown the same observation that Mendel
initially saw, namely that whereas pea plants normally have
purple flowers, one year he finds a pea plant (ostensibly from
last year's purple-flowered parents) with \emph{white} flowers.
We will refer to this trait as \emph{Wh}, and the original (purple-
flowered) trait as \emph{Pu}.

\item {} 
RoboMendel simply seeks to choose the best possible experiment
to do next (mouse X lion?  pea X pea? etc.), performs the experiment,
observes the results, and again chooses the best experiment to do next,
and so on, cycle after cycle.
By ``best'' we simply mean the most efficient path for improving its
prediction power for trait inheritance in general.

\item {} 
We seek a single, general principle for optimizing this process,
in the form of a measure of the likely information yield of an
experiment, expressed in terms of how much it could increase RoboMendel's
prediction power.  That is, we seek a single metric that answers
all of the above questions, ranging from fine details of
experiment design optimization (e.g. how many replicates of this
observation should I collect?  Can I terminate this experiment now
based on the results I've obtained so far?) to big decisions
about what phenomena to study (e.g.
should we study mouse X lion, or pea plant flower color variation?).
Specifically, we mean that it should choose the experimental design
that maximizes this information metric (typically expressed in terms
of information yield per cost, assuming that both ``setup costs'' and
``per observation costs'' are properly accounted for).

\end{itemize}

\subsection{Initial Parameters}
\label{exptplan:initial-parameters}
The RoboMendel problem begins with the observation of one
pea plant with white flowers, apparently descended from regular
purple-flowered parents (since all pea plants previously
encountered by RoboMendel had purple flowers).  We first enumerate
the basic probability factors that frame his experiment planning
process:
\begin{itemize}
\item {} 
\emph{p(LFLS)}: this represents our confidence in our current
like-father-like-son model of genetics.  Note that this reflects
both repeated observations that fit the model (i.e. each time
we actually observe an animal or plant born from its parent,
we see that it looks like the same species as its parent(s)),
and an implicit absence of observations that violate the
model (e.g. we never see evidence of successful inter-species
crosses, like a zebra with a long neck like a giraffe).
Assuming that we have observed on the order of 1000 species,
we might set this probability at \emph{p(LFLS)} = 0.999.

\item {} 
\emph{p(Wh-heritable)}: this estimates the probability that \emph{Wh} is
actually a heritable trait.  Note that if \emph{Wh} is not a heritable
trait (for example, it might be caused by an environmental
factor), then experiments studying it will have zero information
value for our targeted metric, which focuses specifically on
genetic inheritance.  In the absence of any previous
data on this trait, we assume an uninformative prior
\emph{p(Wh-heritable)} = 0.5.

\item {} 
\emph{p(same-species)}: this estimates the probability that \emph{Wh}
is a member of the same species as \emph{Pu}, the regular purple-
flowered pea plant.  Note that \emph{same-species} does not
assume \emph{Wh-heritable}; they are two separate hypotheses.
Again we assume an uninformative prior \emph{p(same-species)} = 0.5.
Intuitively, on the one hand our prior belief was that
it was descended from \emph{Pu} parents, but on the other
hand it doesn't look like them.

\end{itemize}

Note that if both \emph{Wh-heritable} and \emph{same-species} are true,
this would strongly contradict \emph{LFLS}.

\section{Foundations of the Experiment Planning Information Metric}
\label{exptplan:foundations-of-the-experiment-planning-information-metric}
In this section we outline a general information metric for
experiment planning.  This metric is based on standard concepts in
statistical inference and information theory \cite{Shannon48} 
\cite{Kullback51}  \cite{Paninski05}  \cite{Lee2011} ; our interest here is
how they apply to the specific problems of experiment planning and
the scientific method.

\subsection{Empirical Information}
\label{exptplan:empirical-information}

\subsubsection{Defining ``Prediction Power''}
\label{exptplan:defining-prediction-power}
We have previously defined ``empirical information'' estimators based
on two principles: measuring a model's prediction power strictly
for \emph{observable} variables (rather than inferring hidden variables);
estimating this metric from a random sample of observations,
with a Law of Large Numbers (LLN) guarantee of convergence to a classic
information metric such as relative entropy or mutual information
\cite{Lee2011} .  Here we briefly outline their application
to experiment planning.  We define the relative \emph{information} value
of a model in terms of its increase in prediction
power of one likelihood model $\Psi(X)$ relative to
another $\Psi_0(X)$, as the difference in the
expectation log-likelihoods of the observable $X$ under the two
models
\[
\Delta I = E_{\Omega}(\log{\Psi(X)})
- E_{\Omega}(\log{\Psi_0(X)})
\]
where the expectation is taken under the true (but unknown)
distribution of the observable, $\Omega(X)$.  Although this
cannot be computed directly (because $\Omega(X)$ is unknown), it
can be estimated empirically with a LLN convergence
\[
\overline{L_e(\Psi)} = \frac{1}{n}\sum_{i=1}^n{\log{\Psi(X_i)}}
\to E_{\Omega}(\log{\Psi(X)})
\]
as the sample size $n \to \infty$; we refer to
$\overline{L_e(\Psi)}$ as the \emph{empirical log-likelihood}.
We then define the \emph{empirical information}
$\overline{I_e(\Psi)}$ as the estimated increase in prediction
power relative to the marginal distribution of the observable,
$p(X)$:
\[
\overline{I_e(\Psi)}
= \overline{L_e(\Psi)} - \overline{L_e(p)}
\]
Note that this is a pure likelihood metric (i.e. no consideration of
the prior probability of model $\Psi$), similar to the
classical likelihood metrics used for Maximum Likelihood (ML),
the Akaike Information Criterion (AIC) \cite{Akaike74} ,
Bayesian Information Criterion (BIC) \cite{Schwarz78} , and various
log-likelihood ratio tests.  One difference should be emphasized:
LLN convergence requires that $I_e(\Psi)$ be measured
on data $X_i$ that are conditionally independent of the model
$\Psi$ given the true distribution $\Omega$, i.e. on
\emph{test data} rather than \emph{training data}, in marked contrast with
ML, AIC, BIC and many other metrics designed for model-selection
(i.e. choosing a model by maximizing the metric).

\textbf{Example}: after RoboMendel observes white flowers on some of
his pea plants, he can try to improve the prediction power of his
current model (which predicts that pea plants have only purple
flowers) by adjusting it to predict that a certain fraction of
pea flowers are white (Fig. 1C).  He can then measure the empirical
information gain of the new model on a set of observations of
additional flowers.  This validates that the new model greatly
improves prediction power relative to the old model.
Note however that it does not assess whether the new model truly fits the
observations; we address this below with the \emph{potential information}
metric.

\subsubsection{A Zero Information Principle}
\label{exptplan:a-zero-information-principle}
It is useful to highlight a simple but important feature of the metric:
since information is defined as increased prediction
power, any step that does \emph{not} alter the prediction
(i.e. $\Psi=\Psi_0$) must have zero information value.

This simple idea implies an interesting conundrum.  Consider any
new theory $\Psi'$ proposed in response
to observations that ``do not fit''
the existing model $\Psi_0$.
Replacing the existing model with the new
theory produces positive information (since the new theory
fits the observations better).  Now consider an
experimental test of further predictions that the new model makes.
If we were to perform that experiment and obtain results that
do not fit the model $\Psi'$,
this would presumably have information value
(again, because it would force us to change our predictive model).
If on the other hand we obtained results that fit the model $\Psi'$,
that would appear to have zero information value (because no
change in our predictions would result).  On this basis,
any kind of boolean rule for choosing our ``current model''
(i.e. the ``weight'' assigned to $\Psi'$ must be either 0 or 1)
would appear to lead to the following paradox.  If we assign
$\Psi'$ a weight of 1, then the initial ``proposal'' step
produces positive information, but the expectation information
yield from the experiment to test it would be zero (because the boolean
rule assigned zero probability to the negative result, and the
positive result produces zero information yield).
On the other hand if we assign it a weight of zero, then both the
``proposal'' and ``test'' steps would produce zero information.
This suggests that only a ``fuzzy rule'' that allows it to be
assigned a weight between 0 and 1 can ascribe positive information
value both to proposing a new model \emph{and} testing its predictions
experimentally.

\subsection{Expectation Information and the Hidden Mixture Problem}
\label{exptplan:expectation-information-and-the-hidden-mixture-problem}
How does this principle affect our modeling of observables?
First of all, it introduces a subtle change in the joint
probability of multiple observations.  For example, say RoboMendel
observes a series of flowers from a randomly chosen
individual.  If the individual is drawn from a homogeneous
population (under the LFLS model, from a single species)
then these flower observations are
``independent and identically distributed'' (i.i.d.).
However, if the individual is drawn from a hidden mixture
of different species (e.g. one species with white flowers
and one species with yellow flowers), then the flower observations
are no longer independent; for example, if the first
flower observed is yellow, that greatly increases the
probability that subsequent flowers will be yellow.
Note that the ``fuzzy rule'' principle of introducing a new model with a
fractional weight implies exactly this kind of hidden
mixture and consequent lack of independence.

\subsubsection{The i.i.d. vs de Finetti Exchangeability Problem}
\label{exptplan:the-i-i-d-vs-de-finetti-exchangeability-problem}
This statistical distinction may at first seem subtle, but has fundamental
importance, and deserves clear definition.
Random variables $X_1,X_2,...$
are said to be ``independent and identically distributed'' (i.i.d.)
if their joint distribution can simply be factored into
a product of identical terms $p(X_1,X_2,...)=\prod_i{f(X_i)}$
where the same marginal probability distribution $f()$
applies to each of the random variables.  By contrast,
random variables $X_1,X_2,...$ are said to be ``exchangeable''
if for any length $n$ and index sequences
$i_1,i_2,...i_n$ and $j_1,j_2,...j_n$ (where each $i$
value is unique, and each $j$ value is unique),
the joint probability distributions of the indexed variable
sequences are identical:
$p(X_{i_1},X_{i_2},...X_{i_n})=p(X_{j_1},X_{j_2},...X_{j_n})$
\cite{deFinetti37} .
Clearly, if $X_1,X_2,...$ are i.i.d., then they are also
exchangeable, but the converse is not true.  Consider the
case where the joint probability is
\[
p(X_1,X_2,...)=\sum_{\Omega}{p(\Omega)\prod_i{\Omega(X_i)}}
\]
where the hidden variable $\Omega$ is allowed two or more
possible values $\omega_1,\omega_2,...$.  This expression
is completely invariant with respect to exchange of different
$X_i,X_j$, so the random variables are exchangeable.
However, they are \emph{not} independent; instead they are \emph{conditionally}
independent given $\Omega$.

This definition is associated with the mathematician de Finetti,
who proved that any infinite sequence of exchangeable Bernoulli
variables is equivalent to a mixture of multiple i.i.d. Bernoulli
sequences of the form given by the equation above \cite{deFinetti37} .
In statistics,
this definition occasions some subtle distinctions; that is, out of
the many ``standard results'' that assume i.i.d. random variables,
some of them extend without difficulty to the case where
the random variables are \emph{exchangeable}, while others do not.

De Finetti exchangeability is fundamental to the scientific method
because science faces precisely this problem of an unknown mixture
of hidden distributions.
Canonical descriptions of the scientific method tend to gloss
over this mixture problem, in that they discuss hypothesis testing
as if the hypotheses were either ``true'' or ``false'', i.e.
true every single time we test them or false every single
time we test them.  This may seem like a natural attitude
if one assumes that science only studies ``eternal laws'' i.e.
hypotheses that by definition must be \emph{always true} or \emph{always false} in any
given universe.  But \emph{a priori}, it is invalid to assume that because
a hypothesis (``flowers are yellow'') tested true in one experiment,
that it will therefore test true in all experiments.

Instead, we assert
this should be treated as an \emph{empirical} question.  Specifically,
we only assume the general model of a de Finetti mixture, and then
use multiple observations to infer the weight parameters of the
de Finetti mixture.  This corresponds to replacing a boolean
rule for choosing our current model, with a mixture model
that combines multiple models with probabilities between
zero and one.  Of course, if multiple experiments
in a variety of independent cases all confirm a given hypothesis,
its mixture weight will converge to 1.
As we shall show below, it is precisely this combination
of initial uncertainty about the mixture followed by gradual
convergence to an unchanging mixture, that drives the experiment
planning process by predicting large expectation information
yields for ``new questions'' vs. small information yields for
``old questions'' that have already been tested to convergence.

\subsubsection{Expectation Empirical Information}
\label{exptplan:expectation-empirical-information}
This suggests an obvious extension of the empirical information
concept.  Empirical information represents a
measure of our ability to predict various observables in
a collection of scenarios that we have actually sampled.
From the point of view of the de Finetti mixture, this metric
is incomplete, in that it considers only scenarios we've
already sampled, as opposed to the probable mix of familiar
plus novel scenarios that an unbiased estimator should be
based on.  Conceptually, we formulate this as an expectation
value for the prediction power, taken over our posterior
estimate of the de Finetti mixture.  Intuitively, the value
of this metric arises from two sources.  First, truly general
models have \emph{extrapolation power}; that is, they extrapolate
successfully to regions of observation space that were not
actually sampled in the training process.  Second, the de Finetti mixture
estimation process can convince us through repeated testing
that a model actually has such extrapolation power; that is,
its prediction power for novel cases has been found to be
consistently the same as for the original cases.

\textbf{Example}: RoboMendel's LFLS model predicts what progeny will
look like from any pair of parents, and repeated confirmations of
its predictions increase RoboMendel's posterior estimate of
the fraction of cases that it applies to in the real world,
converging to 100\%.

\subsubsection{Universal vs. Targeted Information Metrics}
\label{exptplan:universal-vs-targeted-information-metrics}
We can either define this metric across all possible observables,
or for a \emph{targeted} set of types of observable outcomes that we
particularly care about.  A \emph{universal} information metric treats
all observables as ``equally valuable''; that is, 1 bit of improved
prediction power for one observable (e.g. flower color) counts
exactly the same in the metric as 1 bit of improved prediction
power for any other observable (e.g. identifying the tumor
type of a patient's cancer biopsy).  By contrast, a \emph{targeted} metric
specifies a particular set of observables that we wish to
be able to predict as accurately as possible.  For example,
for the RoboMendel problem, we restrict the information metric specifically
to predicting traits that actually depend on parental traits
(i.e. they are genetic, not due to some other, environmental
factor).

\subsection{The Experiment Planning Information Metric}
\label{exptplan:the-experiment-planning-information-metric}

\subsubsection{Defining Potential Information}
\label{exptplan:defining-potential-information}
In information theory, the entropy $H(X)$ of a random variable
represents a fundamental bound on its predictability.  For
example, no encoding of a sequence of outcomes $X$ can
achieve an average message length of less than $H(X)$.
Similarly, no model $\Psi$ can achieve an average log-likelihood
$E(\log{\Psi(X)}) > -H(X)$.  This sets a hard limit on
the maximum increase in prediction power we could achieve
by \emph{any} model, relative to the current model.  We define
this difference as the \emph{potential information} \cite{Lee2011} 
\[
I_p(X|\Psi) = -E(\log{\Psi(X)}) -H(X) = D(\Omega(X)||\Psi(X)) \ge 0
\]
Note that by improving our model we could convert some or all
of this potential information $I_p$ to empirical information
$I_e$, but \emph{never} more than this amount.  It is for
this reason we refer to it as the ``potential information''
remaining in the observable $X$.

But how can we actually calculate this metric?  It should
first be emphasized that the basic definition of $I_p$
is equivalent to the relative entropy $D(\Omega(X)||\Psi(X))$
of the true, unknown distribution $\Omega(X)$ vs.
the model $\Psi(X)$.  However, we cannot compute this relative entropy,
since we do not know $\Omega(X)$.  Once again, if we have
a sample of observations $X_i$, we can estimate it
empirically, with a Law of Large Numbers convergence guarantee
\cite{Lee2011} :
\[
\overline{I_p(X|\Psi)} = -\overline{L_e(X|\Psi)} - \overline{H_e(X)}
\]
where $\overline{H_e(X)}$ is the \emph{empirical entropy}.

This has immediate significance for experiment planning.
If we consider two different observables $X,Y$ we
could observe experimentally, the maximum increase in prediction
power that could result from these two different experiments
(by the best possible modeling)
is simply $I_p(X|\Psi)$ vs. $I_p(Y|\Psi)$,
where $\Psi$ is our current model.  Thus in principle
we should choose the experiment that maximizes the
potential information.

\subsubsection{Example}
\label{exptplan:example}
In this and subsequent examples, information metrics were
computed by applying RoboMendel models (described in the text)
vs. random sample data generated from genetics simulations.
All code for generating these results is available at
{https://github.com/cjlee112/darwin} .

RoboMendel has been growing pea plants for several years
(Fig. 1).  They have
always had purple flowers.  He looks into his field and sees purple flowers.
This yields mean and lower bound $I_p$ estimators of -0.003 and -0.032:
nothing of interest.  A white flower enters his field of vision: the mean
estimator rises to +0.036 bits.  Moreover, potential information is localizable
in observation space (because it can be interpreted as a relative entropy
of one spatial distribution vs. another).  In this case, the positive $I_p$
signal can be tracked to a single observation (the white flower) with +83 bits
of $I_p$ (a white flower is wildly unlikely under our current pea-species
model).  Of course, because we only have a single observation, the $I_p$
lower bound estimator remains not significant.  This illustrates the first role
of $I_p$ in experiment planning: when there is a large divergence
between the mean $I_p$ estimator and the lower bound, it indicates a likely
opportunity to produce a large amount of potential information by taking
more observations of the item in question, to raise the lower bound.
Since this prospective $I_p$
yield is the largest he currently has available, RoboMendel looks at 100 more flowers
from this plant and obtains +68 bits of potential information (lower bound),
because all the flowers on the plant turned out to be white.

\begin{figure}[]
\begin{center}
\includegraphics[width=6.5in]{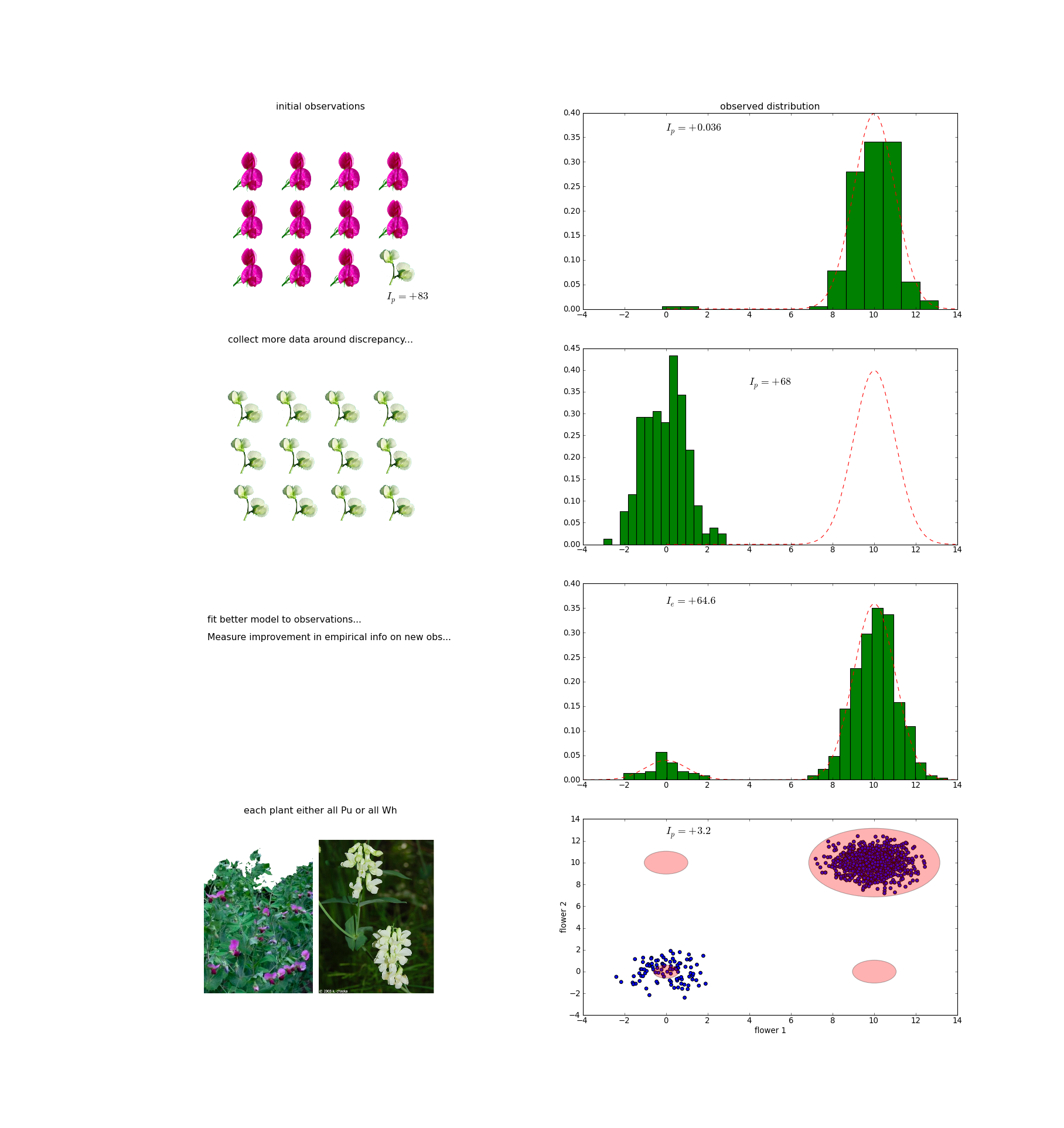}
\end{center}
\caption{
{\bf \textbf{Data collection and model refinement guided by potential information} } 
\textbf{A}. \emph{Detection of a single outlier datapoint is insufficient to raise
the confident lower-bound estimator of} $I_p$, \emph{but suggests that collecting
more observations in the region of the surprising data point could yield
very large potential information.}
\textbf{B}. \emph{When Mendel observes more flowers from this plant, he finds that
all of them are white.  This produces +68 bits of potential information.}
\textbf{C}. \emph{To attempt to convert this potential information to empirical
information, Mendel updates the pea-species model to reflect the ratio
of white vs. purple flowers observed so far.  This converts most of the
potential information to empirical information.}
\textbf{D}. \emph{However, this species model treats each pea plant as drawn
from the same distribution (the pea-species model).  But instead
each plant appears to have one of two different ``phenotypes'': either
all its flowers are purple, or all white.  This mismatch vs. the model
is reflected in strong positive potential information.}
}

\label{exptplan:compositefig}
\end{figure}

Now a second phase begins: seeking to convert this potential information
into new empirical information, by adding new model terms that may fit it better.
Typically the first step of this is to try to use the current model framework
to fit as well as possible.  Let's say RoboMendel has only a simple model of species,
in which each species represents a cluster of observable traits
(e.g. flower color; shape etc.), and each individual is simply drawn from
the probability distribution for its species.  In other words, the only
variable associated with an individual is what species it appears to belong to,
and the only place we can try to tune to ``fit'' the observations better is
the pea-plant species model itself.  In this case we could do that by modifying it
so that the probability distribution for flower color adds a small peak
for white flowers (e.g. representing 10\% of the total density, if that's the
fraction of our pea flower observations to date that have been white).
We then test this new model by taking a new set of observations and calculating
the empirical information yield (relative to our old model).  RoboMendel does this
and obtains +64.6 bits of empirical information.  Is he done?

No, the $I_p$ calculation on the new data still indicates +3.2 bits
of potential information.  The new ``mixture model'' treated each flower
as an independent event (purple or white), whereas in the real observations
we see that the white flowers are segregated to just two individual plants
(that are all-white), while the remaining plants are all-purple.
The lower bound $I_p$ estimator indicates that this is not
an artifact of insufficient sampling; collecting a larger sample
will not make it go away.  Instead, it indicates a convincing
failure of the model.  Note that the $I_p$
calculation did \emph{not} need to be programmed to look specifically for this kind
of hidden order; it was simply detected automatically as a reduction in
the empirical entropy.

\subsubsection{Model Evolution: Prototype, Superset and Posterior Likelihood}
\label{exptplan:model-evolution-prototype-superset-and-posterior-likelihood}
To understand this process of model evolution in a
general way, we need clear terminology
for distinguishing several of its aspects.  First, we refer to
the set of all possible models that are consistent with
past observations as the \emph{model superset}.  Concretely,
every model $\Psi$ with $I_p(\Psi|obs) < \epsilon$
(for some negligibly small value of $\epsilon$)
is a member of this superset.  Thus the only constraint on
such models is that they yield the same likelihood
distribution for the observables collected so far
(they can of course differ not only in their hidden
parameters but also in their likelihood distributions
for \emph{other} observables not yet collected).
Of course, this set is a purely
abstract construct, in that its members are innumerable
(there is no limit to the complexity of models we could propose).

Instead of trying to explicitly enumerate the model superset, modelers
typically propose a single specific model that fits the data.
We will refer to such a model as a \emph{prototype} model; by
definition it is a member of the model superset.
The value of a good prototype model is easily summarized:
\begin{itemize}
\item {} 
modelers generally seek the \emph{simplest} model that fits the
data (Occam's razor).

\item {} 
This simplicity makes the model \emph{predictive}; that is, it
generally asserts a simple \emph{sufficient statistic} that it
claims is the only variable that matters; all other variables
are asserted to be conditionally independent of the observations
given this sufficient statistic.  This claim is a prediction;
that is, it predicts a likelihood distribution for what
we should observe under a variety of different experiments.
To the extent that this likelihood distribution outperforms
the empirical distribution estimated directly from the
raw observations,
we refer to this prediction power as \emph{model information}
$I_m$ \cite{Lee2011} .

\end{itemize}

What happens when we detect strong potential information in
a new set of observations vs. our current favored prototype model?
First, this means that our model superset shrinks, to no longer
include that prototype.  We can of course construct a new
model to fit these observations.  To demonstrate that this
new model actually improves prediction power ($I_e$)
requires collecting a new, independent dataset
(testing the model on the same dataset that was used to
select the model would violate the definition of $I_e$).
This should involve both independent replicates of the original
experiment (this tests the model's predictive power for
\emph{interpolation}), and completely new experiments where the model
predicts a divergent likelihood than the previous model
(we refer to this as \emph{extrapolation}).

How to compute the expected information yield from such experiments?
The de Finetti mixture concept is crucial to this calculation;
that is, we must regard the current probability
of the new prototype model as significantly less than 1,
for obvious reasons:
\begin{itemize}
\item {} 
the new prototype model was constructed by \emph{fitting} the data --
it has not yet been \emph{validated} by an independent test dataset;

\item {} 
it must be considered to be only a subset of current model
superset, i.e. we could easily make up many other models that
also fit the data.  Assuming that some probability measure
exists on this model superset space, the new prototype model's
prior probability should certainly be less than 1.

\item {} 
note that if we asserted a probability of 1 for the model,
any validation experiment would by definition have zero
expectation information value (because successful validation
could not change our predictions at all).

\end{itemize}

Intuitively, the de Finetti mixture probability for
the model $p(\psi_1)$ determines the information yield
of new experiments as follows:
\begin{itemize}
\item {} 
we construct a weighted likelihood
$\psi_{PL}=p(\psi_1)\psi_1+(1-p(\psi_1))\psi_0$, where
$\psi_0$ represents the most plausible ``counter-model''
to the proposed prototype; typically it is just the original
prototype model.  Thus the contrast between $\psi_0$
vs. $\psi_1$ indicates specifically how the new model
makes new predictions.   (Of course, this can be generalized
to more than two competing models $\psi_0,\psi_1,\psi_2...$ etc.)

\item {} 
because the weight $p(\psi_1)$ will change in response
to each new observation, we refer to the weighted likelihood
function $\psi_{PL}$ as the \emph{posterior likelihood}.

\item {} 
successful validation experiments (i.e. that confirm the
new model) drive up $p(\psi_1) \to 1$ and thus change
our overall prediction $\psi_{PL}$, producing
positive information value.

\end{itemize}

\subsubsection{Model Mixture Weighting Schemes}
\label{exptplan:model-mixture-weighting-schemes}
We summarize three distinct schemes for choosing mixture weights
in such de Finetti mixture models:
\begin{enumerate}
\item {} 
\emph{Empirical posteriors}: in cases where there is a historical
record of multiple observations that are relevant to the mixture,
the mixture weights are estimated as a posterior probability
distribution from those observations.  Example: Say RoboMendel
has recorded a set of observations of different individual
animals that allow him to identify each one's species.  Next
he discovers a new species.  Based on all his observations,
he can estimate the frequency of his new species.

\item {} 
\emph{A new distinction}: if the previous observations do not distinguish
the old vs. new models, we may use an uninformative prior.
Example: RoboMendel has observed progeny from many matings of
pairs of a given species (the progeny looked like the same species
as their parents, as predicted by the LFLS model).
Now he observes $Wh \times Pu \to Pu$, which does not fit
LFLS.  So he proposes a new model that fits this observation
(e.g. if the ``father'' was \emph{Pu} and the ``mother'' was \emph{Wh}, he
could propose that only the \emph{father} determines the child's
traits).  Note that in all previous observations, since the
parents were from the same species, \emph{both} models make the same
prediction.  Thus the previous observations provide no basis
for estimating the frequencies of the new vs. old models.
Only the $Wh \times Pu$ experiment distinguishes them,
but since the ``father-only'' model was \emph{constructed} to fit
the results of this experiment, we can hardly treat it as
evidence that any other distinguishing experiment is 100\%
guaranteed to support ``father-only'' too.  Instead, it seems
safer to adopt the conservative position of an uninformative
prior, assigning equal probability to the two models.
We will often use this weighting scheme in the RoboMendel
experiments later in this paper.

\item {} 
\emph{An arbitrary proposal}: say we propose a new model without any
basis in potential information, i.e. the existing model
already fits the observations, so the new model cannot
improve the fit.  This becomes a question
of the \emph{prior} probabilities of the models, which strongly
favor simpler models (Occam's Razor, based on the fact that the number
of possible models goes up exponentially with their complexity, and
their prior probability must be normalized over all such models).
We will not use such arbitrary models in this paper.

\end{enumerate}

\subsubsection{The Expectation Potential Information Metric}
\label{exptplan:the-expectation-potential-information-metric}
We formalize our experiment planning metric in terms of the
expectation value of the potential information of
a specific experiment, under our current estimate of
the probability of different possible ``true distributions''
$\omega_1,\omega_2,...$ (which we may consider to be
the different possible ``hidden states'' of a random variable
$\Omega$ representing the ``true distribution''):
\[
E(I_p(X|\Psi)) = \sum_i{p(\omega_i|\Psi)D(\omega_i(X)||\Psi(X))}
\]
where $p(\omega_i|\Psi)$ represents the probability estimate
under our current model $\Psi$ that the true distribution
of $X$ will turn out to be $\omega_i(X)$.

It is instructive to consider the case where these probability
estimates converge to their true values,
i.e. $p(\omega_i|\Psi) \to p(\omega_i)$:
\[
E(I_p(X|\Psi)) \to \sum_i{p(\omega_i)
\sum_X{\omega_i(X)\log{\frac{\omega_i(X)}{\Psi(X)}}}}
\]\[
= \sum_{i,x}{p(\omega_i,X)\log{\frac{\omega_i(X)}
{\sum_i{p(\omega_i|\Psi)\omega_i(X)}}}}
\]
But the denominator is simply the marginal probability
$\sum_i{p(\omega_i)\omega_i(X)}=p(X)$.  So
\[
E(I_p(X|\Psi)) \to
\sum_{i,x}{p(\omega_i,X)\log{\frac{\omega_i(X)p(\omega_i)}
{p(X)p(\omega_i)}}}
= \sum_{i,x}{p(\omega_i,X)\log{\frac{p(X,\omega_i)}
{p(X)p(\omega_i)}}}
= I(X,\Omega)
\]
i.e. the mutual information of the observable $X$ and
the hidden variable $\Omega$ representing the true,
unknown distribution.  The mutual information is a fundamental
measure of the ``informativeness'' of the observable for distinguishing
the hidden states, and has been proposed as the metric
for choosing an optimal experiment design \cite{Lindley56}  \cite{Machens02} 
\cite{Paninski05}  \cite{Paninski09}  \cite{Cavagnaro10} .
The obvious problem is that the true distribution $p(\omega_i)$
is unknown, and thus we cannot compute the mutual information
as traditionally defined.  However, we can compute the
expectation potential information based on our \emph{subjective}
estimates of $p(\omega_i|\Psi)$.  $E(I_p(X|\Psi))$
represents our subjective estimate of the likely information value
of experiment $X$ given our current model $\Psi$.
In the limit where our subjective
estimates of $p(\omega_i|\Psi)$ converge to the true
values $p(\omega_i)$, then our subjective estimator
of the experiment's value converges to the classical
``objective'' measure given by the mutual information
$E(I_p(X|\Psi)) \to I(X;\Omega)$.  We discuss this
connection further in the Conclusions.

\subsubsection{Disambiguation}
\label{exptplan:disambiguation}
If the observable $X$ can determine the hidden state
of $\Omega$ unambiguously, i.e. if there is no overlap
between the different likelihood distributions $\omega_i(X)$,
then this further reduces to an even simpler case that
we will refer to as \emph{disambiguation}.  Since
\[
I(X;\Omega) = H(\Omega) - H(\Omega|X)
\]
if $H(\Omega|X) \to 0$ (i.e. observing $X$ determines
$\Omega$ unambiguously), then $I(X;\Omega) \to H(\Omega)$.
In other words, the information value of the experiment simply
becomes equal to our initial uncertainty about $\Omega$.
Since good experimental design generally strives to attain an
unambiguous determination, disambiguation is a common scenario.
This also makes mathematically explicit why
repeating an experiment rapidly reduces its
information value; the initial experiment eliminates or greatly diminishes
the uncertainty and thus the information value of repeating the
experiment.

\subsubsection{Example: Mouse x Lion}
\label{exptplan:example-mouse-x-lion}
This experiment directly tests the LFLS model, which predicts
that no offspring should result from an inter-species cross.
The observable ``result'' from the experiment is whether the cross
produces viable progeny or not.  We assume that the possible
observable states \emph{progeny} vs. \emph{no-progeny} can be unambiguously
distinguished.  The probability estimates
of our current model are $\Psi(\text{progeny})=1-p(\text{LFLS})$,
$\Psi(\text{no-progeny})=p(\text{LFLS})$.
To keep things simple, we consider two possible outcomes of
the experiment: if \emph{LFLS} is true, then we will observe
$p(\text{no-progeny})=1$ (i.e. this cross will never produce
progeny); if \emph{LFLS} is false, we will observe $p(\text{progeny})=1$
(every mating successfully produces progeny).  So our
expectation potential information is:
\[
E(I_p) = p(\text{LFLS})D(p(X|\text{LFLS})||\Psi)
+ (1-p(\text{LFLS}))D(p(X|\text{not-LFLS})||\Psi)
\]\[
= p(\text{LFLS}) \left( 1 \log{\frac{1}{p(\text{LFLS})}}\right)
+ (1-p(\text{LFLS})) \left( 1 \log{\frac{1}{1-p(\text{LFLS})}}\right)
\]\[
= H(LFLS)
\]
This simply reflects our assumption that the observable (\emph{progeny} vs.
\emph{no-progeny}) can unambiguously determine the hidden state
(\emph{LFLS} vs. \emph{not-LFLS}).  As we noted above, this reduces the
expectation $I_p$ to a simple disambiguation, where the
experiment's information yield is simply equal to the initial uncertainty
about the hidden state.

For \emph{p(LFLS)} = 0.999, $E(I_p) \approx 0.01$ bits.  The obvious
point is that since our current belief in \emph{LFLS} is already strong,
an experiment testing it yet-again is expected to have low
information value.

\subsubsection{Example: Wh x Wh}
\label{exptplan:example-wh-x-wh}
This experiment tests whether \emph{Wh} is heritable; under the LFLS
model, if \emph{Wh} is heritable (i.e. a distinct species), its
children will also be \emph{Wh}.  To test this prediction,
RoboMendel can simply cross a
white-flowered plant with itself, grow the resulting seeds, and
observe the resulting flower colors.  If he were confident that
\emph{Wh} was definitely a distinct species, this would just be a test of
LFLS and would have a low expectation $I_p$ yield
like the previous case.  However, his current belief
\emph{p(same-species)} = 0.5 introduces uncertainty into what we
expect to see in this experiment.  Specifically,
if \emph{Wh-heritable} then he expects \emph{Wh} progeny; on the other hand
if \emph{same-species} he expects \emph{Pu} progeny.  So as a simple initial
model he adopts $\Psi(Wh)=\Psi(Pu)=0.5$.  The expectation
information yield is
\[
E(I_p) = p(\text{Wh-heritable})D(p(X|\text{Wh-heritable}) || \Psi)
+ (1-p(\text{Wh-heritable}))D(p(X|\text{not-Wh-heritable}) || \Psi)
\]
However, the definition of his targeted information metric
excludes the second term (which represents the case where \emph{Wh}
is \emph{not} heritable), so the yield is reduced to the first term:
\[
E(I_p) = - p(\text{Wh-heritable}) \log{p(\text{Wh-heritable})}
\]
Under his initial assumptions, this gives an expectation yield
of 0.5 bits.  Clearly RoboMendel would prefer this experiment
over \emph{Mouse x Lion}.

\subsubsection{Distinguishing an Experiment's Information Rate vs. Total Capacity}
\label{exptplan:distinguishing-an-experiment-s-information-rate-vs-total-capacity}
When the competing models overlap significantly in their
likelihood distributions,
a single observation is not sufficient to unambiguously
determine the hidden state.  In this case we can repeat the
experiment until the result becomes unequivocal.  This is a
general principle: we can define the \emph{total information capacity}
of an experiment as the limit of information yield as the
number of replicates of the experiment
$n \to \infty$.  Furthermore, we can
define the growth of the information yield as a function of
increasing $n$ as the experiment's \emph{information
yield curve}.  Finally, we can define the slope of this curve at
a given point as the \emph{information rate}.

In general, factors
that are independently sampled in each replicate of the experiment
affect the information rate, but not the total, whereas factors
that remain constant over different replicates of the experiment
(such as the experiment design itself) affect the total yield.
Note that the information rate has critical importance for
computing the \emph{efficiency} of an experiment defined as
its information yield per unit cost.  This is determined by
the experiment's initial setup cost $c_s$ and its
unit cost per replicate $c_r$.

Concretely, how do we actually calculate information yield curves?
Conceptually, we simply replace the single-observation variable
$X$ with a vector $\vec X^n$ representing $n$
observations of $X$.  For many problems, the $n$
dimensional vector $\vec X^n$ can be represented without
loss of information by its sufficient statistic $T(\vec X^n)$,
which simplifies the computation.  We then compute
$E(I_p(\vec X^n|\Psi))$ or $E(I_p(T(\vec X^n)|\Psi))$
for different values of $n$.

\paragraph{Information Rate Example}
\label{exptplan:information-rate-example}
Consider a simple experiment to test whether
\emph{Wh} and \emph{Pu} belong to ``the same species'', by crossing them
and planting the resulting seeds (if any) to see if they grow
successfully (Fig. 2).
Imagine that there is a 30\% probability that
``bad weather'' will occur during the experiment, blocking
any seeds from growing (regardless of whether they are fertile).
If an experiment yields progeny (seeds grow) the result is
unambiguous (``\emph{Wh} and \emph{Pu} are the same species''), but
if no seeds grow we have uncertainty as to whether \emph{Wh} and
\emph{Pu} are different species, or simply bad weather occurred.
Since each experiment has an independent probability of bad weather,
if we simply repeat the experiment multiple times, we can
reduce this uncertainty to any level we wish.  Thus,
the bad weather factor affects the experiment's information rate,
but not its \emph{total} information capacity.  Next, consider
the effect of adding a control to our experiment that explicitly
tests for bad weather.  For example, we could also plant seeds
of a cross that we know should grow (e.g. $Pu \times Pu$);
if they fail to grow, we know bad weather occurred.  Adding
this control improves the experiment's information rate,
but has no effect on its total information capacity.

\begin{figure}[]
\begin{center}
\includegraphics[width=4in]{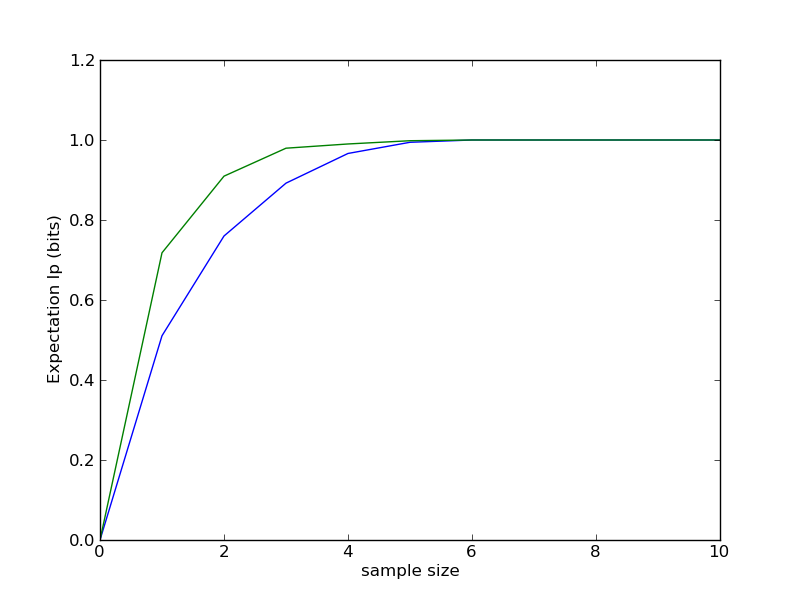}
\end{center}
\caption{
{\bf \textbf{Expectation Ip Yield Curve for Wh x Pu Test} } 
\emph{In the absence of a control (blue) vs. with Pu x Pu control (green).
Assumes 30\% probability of bad weather in each experiment;
50\% probability that Wh and Pu are same species vs. different
species.}
}

\label{exptplan:controlfig}
\end{figure}

\paragraph{Information Total Example: ``Technical Failure''}
\label{exptplan:information-total-example-technical-failure}
In many fields, ``technical problems'' can cause an experiment to fail
to give the predicted result even if the hypothesis is
correct.  For example, in molecular biology, a lengthy
experiment such as knocking out a gene in mouse can simply
give a negative result (no apparent phenotype),
e.g. because another pathway
exists that can complement the target gene function.
Indeed, in complex, incompletely characterized systems such as
biology, any one of a myriad of unpredictable problems
can cause an experiment to fail: e.g. the sample is lacking a factor
that is crucial for the desired reaction; the sample is
insufficiently pure and is contaminated with an inhibitor
that blocks the reaction; the sample is ``too old'' and has
been degraded by the action of proteases or nucleases; etc.
Typically, such technical problems cause a ``negative result'' that
looks the same as would be expected if the hypothesis were false.
As the probability of such technical problems increases,
the expectation information yield decreases for two reasons:
first, the probability of an unambiguous ``confirmation''
(the observed outcome matches that predicted by our hypothesis)
decreases; second, the ``failure'' outcome (the observed result
does not fit the prediction) becomes more and more ambiguous,
i.e. it may not mean that the hypothesis is wrong, it may simply
mean that a technical problem occurred.

Such ``technical failures'' affect the total $E(I_p)$ yield
(rather than just the rate), because the cause of the failure
is built into the experiment design itself.  Simply repeating
the same experiment many times will \emph{not} allow us to sample
over independent draws of \emph{failure} vs. \emph{no-failure}; if the design
is flawed, then \emph{every} replicate will fail in the same way.
Say we are testing a boolean hypothesis $\Psi$ with prior probability
$p(\psi^+)=\alpha$ using an experiment design with
failure probability $f$. If the experiment
yields a boolean observation $x^+$ vs. $x^-$,
then the joint probabilities are $p(\psi^+,x^+)=\alpha(1-f)$
(true positive); $p(\psi^+,x^-)=\alpha f$ (false negative);
$p(\psi^-,x^-)=1-\alpha$ (true negative).
We can compute the total expectation potential information
for this design as the mutual information $I(\Psi;X)$,
which decreases to zero with increasing $f$
(Fig. 3).

\begin{figure}[]
\begin{center}
\includegraphics[width=4in]{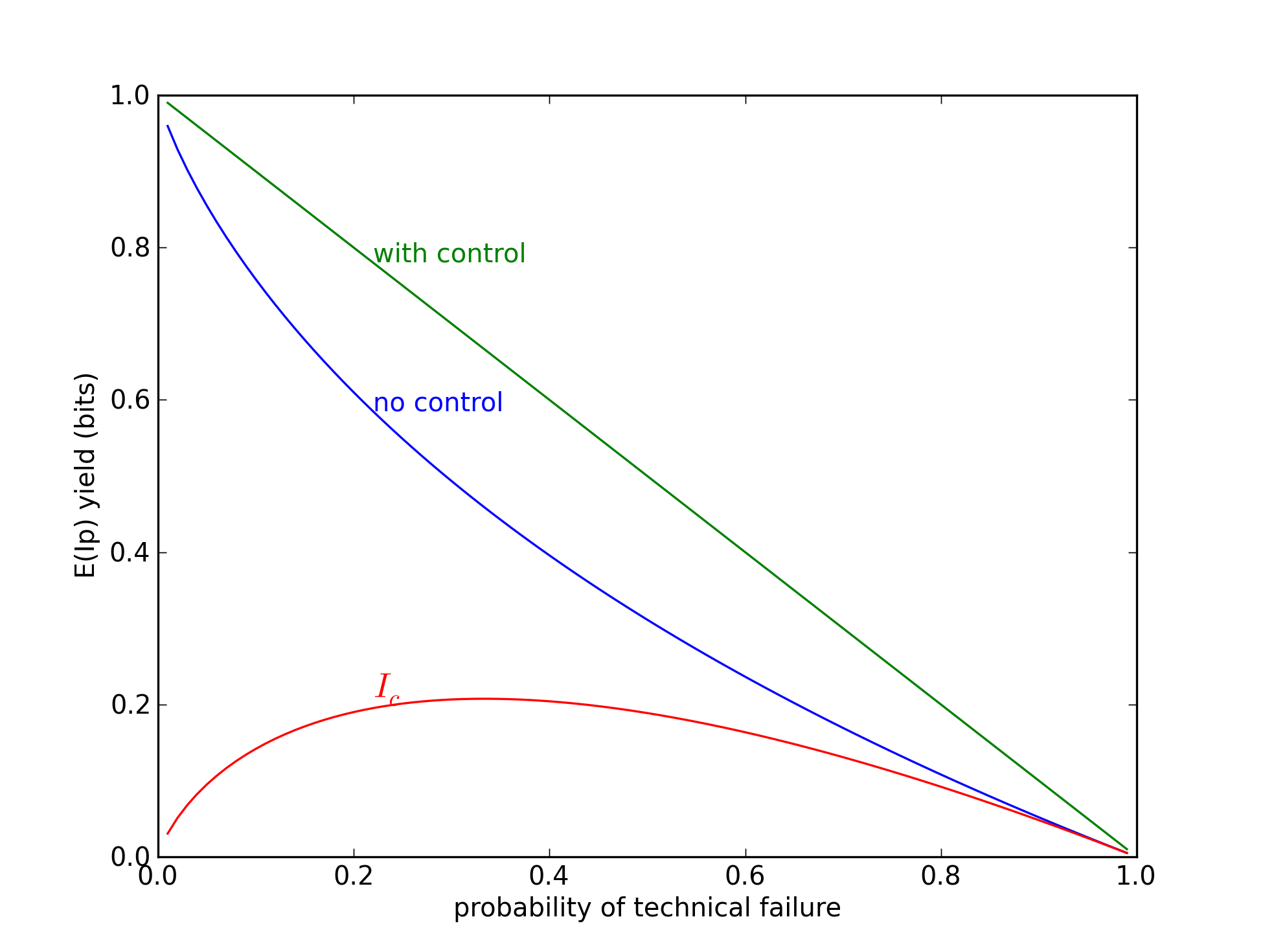}
\end{center}
\caption{
{\bf \textbf{Reduced total expectation potential information as a function of increasing probability of technical failure in an experiment design} } 
$E(I_p)$ \emph{with a positive control (green) vs. without a
positive control (blue); the difference constitutes the}
control information $I_c$
\emph{(red).  Assuming an uninformative prior for the hypothesis}
i.e. $\alpha=0.5$ \emph{(see text for details).}
}

\label{exptplan:envfacfig}
\end{figure}

This analysis also can measure the value of
adding a ``positive control'' to the experiment.  If we add an
observable $C$ that yields value $c^-$ in the case
of technical failure, and $c^+$ otherwise, the joint
probabilities become $p(\psi^+,x^+,c^+)=\alpha(1-f)$
(true positive); $p(\psi^+,x^-,c^-)=\alpha f$ (false negative);
$p(\psi^-,x^-,c^+)=(1-\alpha)(1-f)$ (true negative);
$p(\psi^-,x^-,c^-)=(1-\alpha)f$ (true negative).
This yields a strictly linear decrease in information yield;
thus we can associate with the positive control the amount
of $E(I_p)$ it ``rescues'' (relative to the experiment lacking
the positive control); we refer to this as the \emph{control information}
$I_c$ (Fig. 3).

\subsubsection{Targeted Potential Information}
\label{exptplan:targeted-potential-information}
So far we have ignored the question of whether the observable outcomes are
actually of interest.  In other words, we have presented \emph{universal}
metrics that treat all observables as ``equally valuable'' as targets
for prediction.  It is useful to consider the case where we wish
to target our information metric to a specific set of phenomena.
For example, for the RoboMendel problem, we wish to restrict the
metric to traits that are actually genetic (i.e. which depend
on ancestral traits, as opposed to other, environmental factors).
We accomplish this in the simplest way possible: we just multiply
the expectation information yield for a given observable $X$ by
our probability estimate $\tau_X$ that it matches our target
definition:
\[
E(I_p(X,Y,Z...|\Psi,\tau)) = \tau_X E(I_p(X|\Psi)) +
\tau_Y E(I_p(Y|\Psi)) +\tau_Z E(I_p(Z|\Psi)) + ...
\]
For example, when RoboMendel first encounters a pea plant with
white flowers (\emph{Wh}), there are many potentially interesting questions
he could investigate, but he has fundamental uncertainty
whether \emph{Wh} is even a ``genetic phenomenon'' (i.e. whether
this trait is genetically heritable).  If not, gaining prediction
power for this trait has zero value for his targeted information
metric.  This initial uncertainty has two effects.  First,
it reduces the estimated information yield for the many possible experiments
he could perform on \emph{Wh}, because they are down-weighted by
an initial $\tau_{Wh}=0.5$ (we will use uninformative
priors throughout for such ``initial uncertainty'' values).
Second, it creates strong potential information for any experiment
that can test whether \emph{Wh} is heritable, for the following reason.

\subsubsection{Example: An Environmental Factor Control}
\label{exptplan:example-an-environmental-factor-control}
It is useful to make explicit the factors that could confound this
analysis.  RoboMendel's targeted information metric focuses on
genetically heritable variation, as opposed to variation caused
by environmental factors.  Suppose \emph{Wh} is caused by an environmental
factor such as chemicals in the soil.
That would confound our interpretation of the
\emph{Wh x Wh} experiment; specifically, if we obtain a \emph{Wh} child, we
are uncertain whether that means \emph{Wh-heritable}, or simply that
the environmental effect occurred.  Assume that \emph{p(Wh-env)} is
the probability that any given plant is turned \emph{Wh} by the
unknown environmental factor.  (Clearly we do not expect that
\emph{p(Wh-env)} is 100\%, since most of our pea plants are \emph{Pu}).
Now our model becomes $\Psi(Wh)=(1+p(\text{Wh-env}))/2$.

This reduces our expectation information yield (Fig. 4).
In particular,
it reduces the information rate, but not the total information
capacity of this experiment.  Say we repeat this experiment
$n$ times.  If \emph{Wh-heritable}, we expect to get a \emph{Wh}
child every time.  However, the probability of that outcome
under the environmental factor model is just $p(\text{Wh-env})^n$,
so our combined model becomes
$\Psi(Wh^n)=(1+p(\text{Wh-env})^n)/2$.

\begin{figure}[]
\begin{center}
\includegraphics[width=4in]{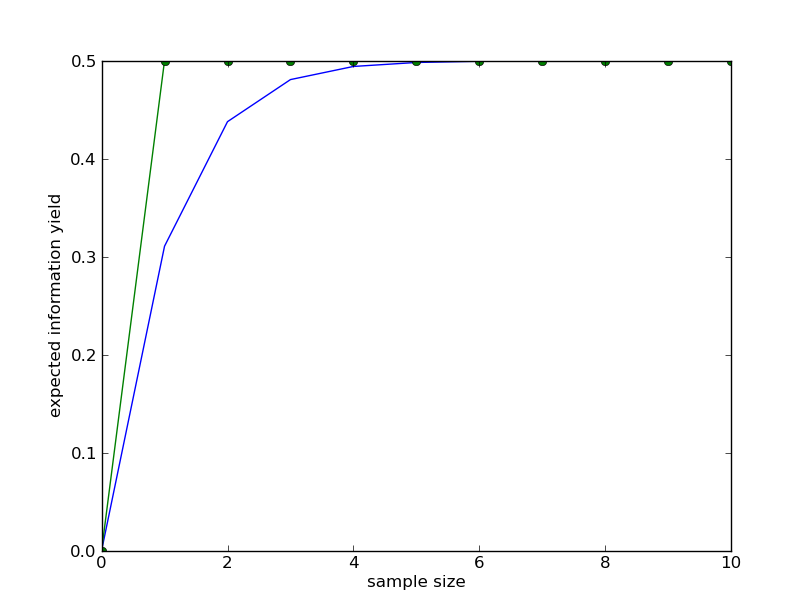}
\end{center}
\caption{
{\bf \textbf{Expectation Ip Yield Curve for Wh x Wh Test} } 
\emph{In the absence of a control (blue) vs. with Pu x Pu control (green).
Assumes 30\% probability of environmental factor in each experiment;
50\% probability that Wh is heritable.}
}

\end{figure}

Next, consider the effect of adding a \emph{Pu x Pu} control, in other
words, planting a seed of a \emph{Pu x Pu} cross immediately next to
the \emph{Wh x Wh} seed.  If \emph{Wh}  is due to an environmental factor
in that patch of soil, then \emph{both} of the resulting plants will
be \emph{Wh}, whereas if \emph{Wh-heritable} then only the \emph{Wh x Wh} plant
will be \emph{Wh}, and the control will be \emph{Pu}.  This control
enables us to unambiguously eliminate the possibility of an
environmental effect even with a single experiment (see figure).

\section{RoboMendel Experiment Planning}
\label{exptplan:robomendel-experiment-planning}
We now apply these basic concepts and metrics to the RoboMendel
experiment planning problem.  To show how the metrics work in practice,
we first apply them to specific experiments.
We then use the metrics to investigate the ``research path'' (the sequence
of experiments chosen by the maximum $E(I_p)$ metric)
under different model assumptions.  Source code for the calculations
for each specific experiment is available from
{https://github.com/cjlee112/darwin}.  Section 5 also provides
a very brief summary of the calculation methods for the
reader to understand them ``at a glance''.

\subsection{A Standard RoboMendel Experiment Set}
\label{exptplan:a-standard-robomendel-experiment-set}
One basic advantage of the RoboMendel problem is that it is
possible to easily enumerate all the possible experiments,
which are simply all possible crosses of various species.
We now list a set of ``standard'' experiments that RoboMendel
will compute expectation information yields for at each stage
of the experiment planning process.  We state
each experiment in terms of the basic cross, and the observed data
obtained if the experiment is performed.  We use the simple
genetics simulation classes included in the \emph{darwin} empirical
information metrics software package \cite{Lee2011} 
to simulate these experiments.
\begin{itemize}
\item {} 
\emph{Mouse x Lion}: described above.  Outcome: no offspring.
Note that this is a generic example of any arbitrary
species-cross experiment; other possible inter-species
crosses would give the same expectation $I_p$ yield.

\item {} 
\emph{Wh x Wh}: described above.  Outcome: \emph{Wh} offspring.  May
include \emph{Pu x Pu} control to test for possible environmental effects.

\item {} 
\emph{Wh x Pu}: outcome, purple-flowered offspring, \emph{Hy} in the nomenclature
below.

\item {} 
\emph{Wh father x Pu mother} vs. \emph{Wh mother x Pu father}: outcome,
purple-flowered offspring.

\item {} 
\emph{Pu father x Pu mother} vs. \emph{Pu mother x Pu father}: outcome,
purple-flowered offspring.  This design simply makes explicitly
obvious that the ``swap'' experiment has no value unless
the mother and father have different traits.

\item {} 
\emph{Pu x Pu self-cross} (cross a \emph{Pu} individual to itself).
Outcome: for each heterozygous locus, 1/4 of the offspring will
be homozygous.

\item {} 
\emph{Hy x Hy}: outcome, 1/4 \emph{Wh}, 1/4 \emph{Pu}, 1/2 \emph{Hy} offspring.

\item {} 
\emph{Wh x Hy}: outcome, 1/2 \emph{Wh}, 1/2 \emph{Hy} offspring.

\item {} 
\emph{Pu x Hy}: outcome, 1/2 \emph{Pu}, 1/2 \emph{Hy} offspring.

\end{itemize}

Of course, it must be emphasized that RoboMendel does not any
know these outcomes, but simply chooses an experiment to perform
(based on the $E(I_p)$ yield for all these experiments),
performs the experiment, and collects observations.
Of course, these data will in turn change the expectation information
yields for all of the experiments (for example by altering
RoboMendel's probability estimates for \emph{p(Wh-heritable)},
\emph{p(same-species)}, \emph{p(LFLS)}, etc.).  RoboMendel therefore
recomputes the $E(I_p)$ yields, chooses the next
best experiment to perform, and the cycle repeats.

\subsection{The RoboMendel Experiment Sequence}
\label{exptplan:the-robomendel-experiment-sequence}

\subsubsection{Experiment 1}
\label{exptplan:experiment-1}
Starting from RoboMendel's initial uninformative parameters,
it computes the following expectation information yields
for the standard experiments:

\begin{table}[h!]
\caption{
\bf{Expectation potential information yields from the initial conditions}}
\begin{tabular}{|l|l|}

\hline
\textbf{
Experiment
} & \textbf{
$E(I_p)$
}\\
\hline

\emph{Wh x Wh}
 & 
0.5 bits
\\

\emph{Wh x Pu}
 & 
0.09 bits
\\

\emph{Mouse x Lion}
 & 
0.01 bits
\\

\emph{Wh x Pu swap}
 & 
$1.2 \times 10^{-6}$ bits
\\

\emph{Pu x Pu swap}
 & 
0 bits
\\

\emph{Pu x Pu self-cross}
 & 
0 bits
\\
\hline

\end{tabular}
\end{table}

\textbf{Comments}: the \emph{Wh x Pu} experiment yields are reduced by strong
uncertainty about both \emph{same-species} and \emph{Wh-heritable}.  The
\emph{Wh x Wh} experiment gives a strong yield, because RoboMendel
is completely unsure what its result will be, and because
it can reveal whether \emph{Wh} is heritable.

RoboMendel chooses to perform the \emph{Wh x Wh} experiment, and obtains
100\% white-flowered progeny.  As the sample size (number of
progeny) $n \to \infty$, this causes
$p(\text{Wh-heritable}) \to 1$.  To keep this discussion
simple, we skip over these details and assume that at the
termination of the experiment $p(\text{Wh-heritable})=0.999$.

\subsubsection{Experiment 2}
\label{exptplan:experiment-2}
This leads to a new set of expectation information yields:

\begin{table}[h!]
\caption{
\bf{Expectation potential information yields after Wh x Wh experiment}}
\begin{tabular}{|l|l|}

\hline
\textbf{
Experiment
} & \textbf{
$E(I_p)$
}\\
\hline

\emph{Wh x Pu}
 & 
0.19 bits
\\

\emph{Mouse x Lion}
 & 
0.01 bits
\\

\emph{Pu x Pu swap}
 & 
0.001 bits
\\

\emph{Wh x Wh}
 & 
0.001 bits
\\

\emph{Pu x Pu self-cross}
 & 
0 bits
\\

\emph{Wh x Pu swap}
 & 
0 bits
\\
\hline

\end{tabular}
\end{table}

\textbf{Comments}: the \emph{Wh x Wh} yield drops near zero because the previous
results convinced us that \emph{Wh} is heritable (so there is no uncertainty
what the outcome of repeating this will be).  The \emph{Wh x Pu} yield
has doubled because our uncertainty about \emph{Wh-heritable} was eliminated.

RoboMendel therefore performs the \emph{Wh x Pu} experiment and obtains
100\% purple-flowered progeny.  The fact that progeny were produced
at all is only consistent with the \emph{same-species} hypothesis.  Again
ignoring the details of sample size, we assume that RoboMendel
performed a large enough experiment to obtain a reasonable degree
of confidence, leaving him with $p(\text{same-species})=0.999$

We will refer to the progeny of this experiment as \emph{Hy} (hybrid),
since they represent a potentially new category (the offspring of
\emph{Wh} and \emph{Pu}, parents that do not ``look like'' they belong to the
same species, at least according to our current pea species model).
This adds some new possible crosses to the list of experiments.

\subsubsection{The One-Parent Model}
\label{exptplan:the-one-parent-model}
The asymmetry of this experimental result (the offspring resemble
one parent but not the other) raises questions about what exactly
the role of both parents is in determining the child's traits.
Previously, such questions did not arise because both parents
were drawn from the same distribution.  One possible explanation
of this result would be if there was a simple asymmetry in which
only \emph{one} parent's traits were actually passed on to the child.
For example, if RoboMendel's \emph{Wh x Pu} experiment crossed
\emph{Wh} mothers (flowers) with \emph{Pu} fathers (pollen), he might
interpret the result as indicating that only the \emph{father} determines
the child's traits.  Note that this is irrelevant to
all previous experiments (where both parents resemble each other).
Therefore RoboMendel proposes this new variant of the LFLS
model, with an uninformative prior of \emph{p(one-parent)=0.5}.

\subsubsection{Experiment 3}
\label{exptplan:experiment-3}
RoboMendel computes a new set of expectation information yields on
this basis:

\begin{table}[h!]
\caption{
\bf{Expectation potential information yields after Wh x Pu experiment}}
\begin{tabular}{|l|l|}

\hline
\textbf{
Experiment
} & \textbf{
$E(I_p)$
}\\
\hline

\emph{Wh x Pu swap}
 & 
1.0 bits
\\

\emph{Mouse x Lion}
 & 
0.01 bits
\\

\emph{Pu x Pu swap}
 & 
0.001 bits
\\

\emph{Wh x Wh}
 & 
0.001 bits
\\

\emph{Pu x Pu self-cross}
 & 
0 bits
\\

\emph{Wh x Pu}
 & 
0 bits
\\
\hline

\end{tabular}
\end{table}

\textbf{Comments}: RoboMendel's uncertainty about whether LFLS might actually
only depend on one parent makes the swap experiment highly informative
(a straightforward disambiguation of a question with 50\% uncertainty).
Note that it receives its full expectation Ip value because RoboMendel
is now confident of both \emph{same-species} and \emph{Wh-heritable}.

RoboMendel therefore performs the swap experiment, and obtains
100\% purple-flowered progeny regardless of which parent is \emph{Wh}
vs. \emph{Pu}.  Assume RoboMendel's sample size is large enough
to reduce the \emph{one-parent} model to \emph{p(one-parent)=0.001}.

\subsubsection{The Transmission Model}
\label{exptplan:the-transmission-model}
RoboMendel therefore considers the next level of model, specifically
a Markov model in which each individual receives a ``signal'' from each
of its parents, and transmits one of these signals to each of its
children.  By definition, \emph{Pu} sends a \emph{pu} signal, and \emph{Wh} a
\emph{wh} signal.  Thus the transmission model predicts that
since \emph{Hy} received 1 \emph{wh} + 1 \emph{pu} signal, it will send
\emph{wh} with 50\% probability and \emph{pu} with 50\% probability to its children.
This implies a necessary corollary that \emph{pu} is dominant
over \emph{wh}, so an individual containing both (like \emph{Hy})
will be purple-flowered.  The conventional terminology for
this is to call \emph{Wh} a \emph{recessive trait}.
Again, RoboMendel assigns this model an uninformative prior
of \emph{p(transmission)=0.5}.

Note that the real significance of this model is that it introduces
a \emph{hidden variable} into our genetic model.  That is, it postulates
that the inheritance of traits is determined by ``signal states''
which are not necessarily directly observable.  Specifically, it
postulates that even though \emph{Hy} is purple-flowered (and looks no
different than \emph{Pu}), it contains a hidden \emph{wh} signal.

\subsubsection{Experiment 4}
\label{exptplan:experiment-4}
RoboMendel computes a new set of expectation information yields on
this basis:

\begin{table}[h!]
\caption{
\bf{Expectation potential information yields after Wh x Pu swap experiment}}
\begin{tabular}{|l|l|}

\hline
\textbf{
Experiment
} & \textbf{
$E(I_p)$
}\\
\hline

\emph{Hy x Wh}
 & 
1 bits
\\

\emph{Hy x Hy}
 & 
0.98 bits
\\

\emph{Mouse x Lion}
 & 
0.01 bits
\\

\emph{Pu x Pu swap}
 & 
0.001 bits
\\

\emph{Wh x Wh}
 & 
0.001 bits
\\

\emph{Pu x Pu self-cross}
 & 
0 bits
\\

\emph{Wh x Pu}
 & 
0 bits
\\

\emph{Wh x Pu swap}
 & 
0 bits
\\

\emph{Hy x Pu}
 & 
0 bits
\\
\hline

\end{tabular}
\end{table}

\textbf{Comments}: both the \emph{Hy x Hy} and \emph{Hy x Wh} experiments receive
high expectation Ip values, because for a reasonable sample size
(e.g. $n=20$) they turn into a straightforward disambiguation
of two distinct predictions (according to LFLS, \emph{Hy} should yield
offspring that resemble it (i.e. 100\% should be purple-flowered);
whereas \emph{transmission} predicts that 25\% of its progeny should
receive two \emph{wh} signals and therefore should be white-flowered).
Note that \emph{Hy x Pu} gives no expectation information, since both
models predict the same observable outcome (100\% purple-flowered
progeny).

RoboMendel chooses to perform the \emph{Hy x Hy} and \emph{Hy x Wh} experiments
simultaneously (they can both be done at the same time).
The results fit the \emph{transmission} model, and reject the \emph{LFLS}
model (at least for the \emph{Wh/Pu} trait).

\subsubsection{The Value of Predictions}
\label{exptplan:the-value-of-predictions}
The \emph{transmission} model has correctly predicted the
behavior of one genetic trait.  As we emphasized earlier in this
paper, this does not prove that it will hold true for \emph{all}
genetic traits.  Instead, this must be viewed as an unknown
de Finetti mixture which we can only learn empirically, i.e.
to test experimentally what fraction of genetic traits obey the
\emph{transmission} model.  Concretely, how can we best do this?

The \emph{transmission} model itself suggests an experiment that can
begin to answer this question.  If any additional traits exist
in the population that obey this model, we should be
able to reveal them easily by a self-cross experiment.
That is, additional recessive traits like \emph{Wh} may be present in the
population but \emph{hidden} because they are infrequent,
and therefore very unlikely to occur in \emph{both} copies
in any individual (which would be necessary
for it to produce a visible effect).  The key prediction
of the \emph{transmission} model is that any self-cross
(in which an individual is mated with itself) will
reveal approximately one-quarter of its recessive traits
in \emph{each} of its children.  That is, just like for \emph{Hy x Hy},
each recessive trait has a 1/4 probability of transmitting
two copies of itself to a given offspring.  Thus the
\emph{transmission} model predicts that the self-cross experiment
is an effective way to see whether any additional
``Mendelian'' traits actually exist in the population.

Once again, RoboMendel's assessment of the value of this
experiment depends on his prior probability for this novel
event.  As before, we assume an uninformative prior, that is
\emph{p(more-traits)=0.5}.  How can RoboMendel calculate an
expectation information yield for ``unknown traits''?  The
key is that the $E(I_p)$ doesn't require predicting
exactly what a novel trait will look like; it only requires an
estimate of its relative entropy vs. RoboMendel's current model
of a \emph{Pu x Pu} cross.  RoboMendel only has one example mutant trait
(\emph{Wh}) to build such an estimate upon.  \emph{Wh} was observed to
be 10 standard deviations divergent from the \emph{Pu} distribution
on one variable (flower-color) but apparently identical to \emph{Pu}
in its distribution of other variables (e.g. plant size; number
of leaves; shape of leaves etc.).  We therefore imagine modeling
a similar ``single-variable divergence'' as follows.  Following
our \emph{p(more-traits)} prior, we assign 50\% of the probability
to an uninformative density on this variable (we reserve the
remaining 50\% for the standard \emph{Pu} peak for this variable).
In order to cover deviations as large as \emph{Wh} in both directions,
the uninformative part of this density is of the form $p(X)=0.5/20$
for $X \in [0,20]$.
Under such a model the relative entropy of a trait like \emph{Wh} would be
$D(Wh||\Psi)=\log40-H(X|Wh) \approx 3.27$ bits.

Therefore the expectation $I_p$ yield for finding one such
trait with probability \emph{p(more-traits)=0.5} is approximately 1.64 bits.
Note this assumes generating enough progeny from a given individual
(e.g. $n=20$) for strong confidence of getting at least
one child with two copies of the putative trait (with 25\% probability
per child).

Note that this expectation yield is a direct consequence of the
de Finetti mixture model.  That is, we \emph{selected} the \emph{transmission}
model to fit a single instance of a genetic trait.  But this single
instance does not prove that the model will be a ``universal law''
that applies to all genetic traits.  Instead, according to the
de Finetti mixture there is an unknown mixture of trait types
(some of which may obey this model, and others which obey
different models).  The associated probabilities have a crucial
effect on the model's total prediction power, depending on
whether most traits obey it vs. very few.
(Taking the \emph{transmission} model as a specific example, even today researchers
struggle with the question of what proportion of important
human disease traits are ``simple'' (involving only a few
mutations / genes; in the simplest case, Mendelian)
vs. ``complex'' (in which a disease susceptibility arises
from a combination of small effects of many genes)).
According to the de Finetti view we can only learn
these mixture probabilities empirically, e.g. by testing different
genetic traits to see if they obey this model.  Based on a single
case where we \emph{selected} the \emph{transmission} model to fit the data,
we have little basis for estimating this mixture, so initially
we use an uninformative prior.

This considerably changes the expectation yield for the self-cross
experiment.  Previously it had zero value, because RoboMendel's current
model confidently expected a single outcome (\emph{Pu} children, just as
would be predicted for any \emph{Pu x Pu} cross).  Now, however, it
provides a powerful way to test a hypothesis that RoboMendel is
currently very uncertain about (i.e. whether any more traits will
obey the \emph{transmission} model).

Another question is how many distinct individuals
RoboMendel should perform the self-cross experiment on; this
depends on the (unknown) frequency of such traits in the population.
Again using \emph{Wh} as a guide, it seems reasonable to use approximately
the same number of individuals in which the \emph{Wh} trait was
first found, say approximately 20 - 100 plants.
In this way, RoboMendel would have the same
power for detecting an additional trait as was used for
discovering the original \emph{Wh} trait.

\subsubsection{Experiment 5}
\label{exptplan:experiment-5}
RoboMendel obtains a new set of expectation information yields on
this basis:

\begin{table}[h!]
\caption{
\bf{Expectation potential information yields after Hy x Hy/Wh experiments}}
\begin{tabular}{|l|l|}

\hline
\textbf{
Experiment
} & \textbf{
$E(I_p)$
}\\
\hline

\emph{Pu x Pu self-cross}
 & 
1.64 bits
\\

\emph{Mouse x Lion}
 & 
0.01 bits
\\

\emph{Pu x Pu swap}
 & 
0.001 bits
\\

\emph{Wh x Wh}
 & 
0.001 bits
\\

\emph{Hy x Hy}
 & 
0.001 bits
\\

\emph{Hy x Wh}
 & 
0.001 bits
\\

\emph{Wh x Pu}
 & 
0 bits
\\

\emph{Wh x Pu swap}
 & 
0 bits
\\

\emph{Hy x Pu}
 & 
0 bits
\\
\hline

\end{tabular}
\end{table}

RoboMendel therefore performs the \emph{Pu x Pu self-cross} experiment
and inspects the progeny for clearly anomalous characteristics.
Based on historical data, this experiment would be highly likely
to discover additional recessive traits such as those found
by Mendel: \emph{Wrinkled} seeds; \emph{White seed coats}; \emph{Yellow seeds};
\emph{Yellow pods}; \emph{Constricted pods}; \emph{Terminal flowers}; \emph{Short} plants; etc.

\subsection{What If RoboMendel Fails to Propose the Transmission Model?}
\label{exptplan:what-if-robomendel-fails-to-propose-the-transmission-model}
The above sequence demonstrates the power of proposing a good
model: it predicts new experiments that lead to further discoveries.
This suggests an obvious question: what would have happened if
RoboMendel had not considered the \emph{transmission} model
after the \emph{Wh x Pu} experiment?  Certainly, alternative models are
possible.

\subsubsection{Alternative Model: ``\emph{Pu} Undilutable''}
\label{exptplan:alternative-model-pu-undilutable}
For example, RoboMendel could simply have proposed
that ``\emph{Pu} always beats \emph{Wh}``, in the sense that \emph{any} child
with a purple-flowered parent and a white-flowered parent
will always have purple flowers.  (Note that this is different
than saying that \emph{Wh} is recessive in the \emph{transmission} model).
We now consider this alternative experiment sequence.

The essence of this model is that \emph{Pu} is ``undilutable'': no matter
how much \emph{Wh} we ``dilute'' it with (i.e. how many times we cross
it with \emph{Wh}), we will still get 100\% \emph{Pu} progeny.  This suggests
an obvious experiment: serially ``dilute'' \emph{Pu} by crossing it
over and over with \emph{Wh}.  If the new \emph{Pu-undilutable} model is
correct, these successive generations will remain 100\% purple-flowered;
otherwise we might expect increasing amounts of \emph{Wh} progeny
to begin appearing.  Following our previous practice
we assign the new model an uninformative prior \emph{p(Pu-undilutable)=0.5}.
Note that we already have the first step in this ``dilution
sequence'': \emph{Hy}, the result of crossing ``pure'' \emph{Pu} to \emph{Wh} once.
Concretely, we would like to complete at least a few ``dilution
cycles'' with large enough sample sizes in each generation
to detect a small fraction of \emph{Wh} children.  Say each cross
produces 30 seeds from which new plants can be grown.  If \emph{Pu} were
a chemical compound, its concentration would be diluted 30-fold
each generation.  If RoboMendel repeats this dilution over 10 generations,
\emph{Pu} will be diluted by a factor of $5.9 \times 10^{14}$,
which would appear to be a rigorous test of the dilution model.
RoboMendel therefore computes the expectation information yield
of this experiment as simple disambiguation of a hypothesis
with 50\% uncertainty, which yields 1 bit.

Thus \emph{Hy x Wh} becomes RoboMendel's next experiment step
(highest expected yield).
It yields an unexpected result: half \emph{Wh} progeny, and half
purple-flowered progeny.  Not only is \emph{Pu-undilutable} immediately
rejected, but this result makes clear that RoboMendel must
hypothesize a ``hidden'' variable representing an individual's
genetic state (since two plants that outwardly looked the same,
\emph{Hy} and \emph{Pu}, behaved completely differently genetically).
Note that introducing a ``hidden variable'' to represent the
genetic state is the essence of the \emph{transmission} model.
Thus, failing to propose the \emph{transmission} model, but
continuing to use information metrics to find the best
test of alternative models, has forced RoboMendel back
towards the \emph{transmission} model.

\subsubsection{Alternative Model: An Inter-Species Hybrid?}
\label{exptplan:alternative-model-an-inter-species-hybrid}
We now consider another alternative explanation of the \emph{Wh x Pu} result.
Mendel and contemporaries were aware that in some cases crosses
of highly similar species (e.g. horse and donkey) could produce
viable progeny (e.g.
$\text{horse} \times \text{donkey} \to \text{mule}$),
but that these progeny were generally sterile (i.e. unable to
reproduce).  Thus RoboMendel could propose an alternative
explanation that the \emph{Wh x Pu} result does not imply
\emph{same-species}, by asserting that \emph{Hy} is a ``hybrid cross''
of two different species.  This model suggests an obvious
test: \emph{Hy x Hy}, to see if progeny are obtained.  Assigning
this \emph{species-hybrid} model an uninformative prior of 0.5,
the \emph{Hy x Hy} experiment becomes a straightforward
disambiguation with a 1 bit expected yield.  RoboMendel
therefore performs this experiment as his next best step.

Even with progeny from a single \emph{Hy x Hy} cross (approximately
30 plants), RoboMendel is almost certain to obtain at least one
\emph{Wh} child (since each child actually has a 25\% chance of being
\emph{Wh}).  This both rejects the \emph{species-hybrid} model and again
compels the concept of a ``hidden variable'' describing genetic
state (since again \emph{Hy} is behaving very differently from \emph{Pu}
genetically, even though it looks exactly the same in appearance).

Once again, using expectation information to pick experiments
that will test the model has overcome a wrong initial choice of
model, and forced RoboMendel back in the direction of the
\emph{transmission} model.

\section{Conclusions}
\label{exptplan:conclusions}
Recently, there has been growing interest in automated
experiment planning in general \cite{King09}  \cite{Guan10} 
and specifically in mutual information for optimal experiment planning
\cite{Machens02}  \cite{Nelson05}  \cite{Paninski05} 
\cite{Paninski09}  \cite{Cavagnaro10} .
The approach presented in this paper is largely compatible
with previous results on mutual information as an
experiment planning metric, although it starts from rather
different foundations.  The mutual information approach focuses
on the question of how informative a given experiment (\emph{observable}
variable $X$) is about a designated
\emph{hidden} variable $\Omega$ (the target of
interest).  By contrast, potential information is defined
strictly in terms of the likelihood of \emph{observable} variables
(measuring our prediction power on these observables,
\emph{not} our ability to infer hidden variables).  It is a
striking result that the expectation potential information
(defined in these purely observable terms)
can converge to the mutual information.
This follows from the de Finetti mixture of competing models
in the $E(I_p)$ calculation; when the estimated mixture weights
converge to the true model probabilities,
the expectation potential information converges to the
true mutual information $E(I_p) \to I(X;\Omega)$.
Of course, in real scientific inference problems,
the true model probabilities are unknown.

Ordinarily, the mutual information metric is applied to
problems where the hidden variable target (``what question
to ask?'') is pre-defined, and the experimental observation
(``how to answer it?'') is varied, to find the best experiment
for determining the value of the hidden variable.
As an example of this distinction, consider the behavior of
the general mutual information metric $I(X;\Omega)$
as we repeat an experimental observation of $X$ again
and again.  By definition, $I(X;\Omega)$ remains
constant after these multiple observations, even though
the past observations may have already told us the
actual value of $\Omega$.  This behavior is not
suitable for helping us decide ``what question to ask?''.
By contrast, the expectation potential information
goes to zero if an observation is repeated in this way,
reflecting the fact that there is nothing more to learn
from repeating this experiment.
In this respect it behaves more like a conditional mutual
information taking into account all previous observations.

In this paper we have addressed \emph{both} types of
challenges together.  Expectation potential information
is defined in terms of the possible gain in prediction power
(ability to predict \emph{observables}) that could result from
performing an experiment.  This definition subsumes
the issues not only of how best to answer a given question,
but also what question will be most valuable to ask
(i.e. will yield the largest possible increase in prediction
power).  Concretely, we have demonstrated the utility of
$E(I_p)$ for choosing the ``best next question'' among
very different experiments
(e.g. \emph{mouse} $\times$ \emph{lion}, $Wh \times Pu$ etc.).
We have presented evidence that the expectation potential
information metric is useful both
for guiding \emph{experiment proposal} (choosing what
experiment to do next, out of all the possibilities),
and for fine-grained \emph{experiment
optimization} (such as deciding whether a given control is worth
including or not).  It
generalizes nicely from single observations to multiple replicate
observations, providing straightforward measures of both
\emph{information rate} and \emph{total information yield}.

We suggest that the main value of an experiment planning metric is
when it acts as a ``gradient'' on the space of all possible
experimental observations.  In other words, it provides a purely
local signal (computable at any given point on an experiment
planning trajectory) that indicates how to ``read'' the largest
increase in average prediction power from this space.  Thus
it breaks down the problem of guessing the hidden structure
of this space, into small steps for reading the patterns in
the data, ideally one dimension at a time.  (Of course, this
signal must be present in the data in the first place; if
the observations were encrypted with a one-time pad, no such
breakdown would be possible).

The RoboMendel problem illustrates this process in action.
The initial detection of a potential information signal
(white flowers on a pea plant descended from purple-flowered
parents) indicates a breakdown of the current model,
and suggests plausible modifications of the model.
Computation of expectation potential information identifies
the best experiments to test these plausible models.
An important feature of such a metric is robustness:
ideally it should point you in the right direction
regardless of what path you take to get there.
For example, in addition to considering the ``canonical''
path (in which RoboMendel proposes the ``transmission'' model
that Mendel himself proposed), we also considered several alternative
paths in which RoboMendel proposed alternative, incorrect
models.  We showed that $E(I_p)$ then identified
experiments to test these models, which produced observations
that rejected these models and directly revealed the ``transmission''
pattern.  This at least illustrates the kind of robustness
that we would hope for from a general information metric.

As such, this metric could be applied to enable RoboMendel to
discover many additional features of genetics.  Note that Experiment 5
(Pu x Pu self-cross) would directly lead to identifying the
next layer of genetics, namely multiple genetic loci controling
multiple traits.  From this point, the exact same $E(I_p)$
experiment planning process
should be able to discover the obvious substructure and superstructure
of these traits, e.g. discovering complementation groups
i.e. ``genes''; discovering chromosomes;
constructing genetic maps; discovering pathways and epistasis etc.

Of course, it must be emphasized that far more is required for
actual automated experiment planning and discovery \cite{King09} .  Here we
have only discussed an information metric for forecasting the
yield, given specific proposed models and experiments.
That entirely ignores the crucial questions of how
to generate model proposals in an automatic and appropriate way,
and robust principles for assigning their de Finetti weights
(prior probabilities).  In the RoboMendel example we have also
largely ignored the question of how to generate experiment
proposals (this clearly must be driven by looking for the
predicted differences between the new vs. old models).
Another major area that we've ignored is the question
of experiment costs.  Presumably, to choose the experiment
with the highest information yield per cost requires
good models of the structure of experiment costs, e.g.
distinguishing setup costs and latency for \emph{beginning}
an experiment vs. the unit cost of adding one more sample
replicate.  We point out these large areas of research
merely to emphasize that the information metric presented here
is but a small piece of the whole puzzle.

\subsection{The Scientific Method}
\label{exptplan:the-scientific-method}
Karl Popper emphasized that a scientific hypothesis must be
testable to be useful \cite{Popper59} .
Thus the value of a hypothesis is that
it makes predictions that both differ from our previous
expectations (otherwise they would be indistinguishable),
and are readily verifiable by available experimental means.

The expectation potential information metric is entirely consistent
with this outlook.  In order for an experiment to produce
potential information, it must have two or more observationally
distinguishable outcomes, and there must be strong uncertainty
about which outcome will actually occur.  Translating this
into Popper's terms, this means that the hypothesis under
test must predict a different outcome than our previous
model, and there must be uncertainty about the hypothesis
(leading to uncertainty about the expected outcome).

The RoboMendel example illustrates how the scientific method
translates into a cycle of steps driven by the basic
information metrics of potential information and
expectation potential information.  We enumerate the basic
steps:
\begin{enumerate}
\item {} 
\emph{a surprising observation}: substantial potential
information is detected (via a confident lower bound
on the $I_p$ metric \cite{Lee2011} ).

\item {} 
\emph{a new model that fits}: a modification to the current
model is proposed, typically by fitting the new, surprising
observation(s).  Note that the new model should \emph{not} differ
from the old model on previous sets of observables (which
the old model fit well).

\item {} 
\emph{a divergent prediction}: the new model can only improve
prediction power if it makes some different predictions
than the old model, and if we have significant uncertainty
about their de Finetti mixture weights.  Actually testing
one of these predictions will have information value to
the extent it can result in changing the de Finetti weights.
We quantify this as the $E(I_p)$ metric.

\end{enumerate}

Since the purpose of the $E(I_p)$ metric is to maximize
the potential information yield, performing the experiment it chose
to test the model's predictions often leads to new potential
information, e.g RoboMendel's tests of the ``one-parent'' model
or the ``Pu undilutable'' model both lead to surprising observations
(that do not fit the model).  Then we resume the cycle again at
step 1.

\subsection{The ``Boolean Logic'' of Expectation Potential Information}
\label{exptplan:the-boolean-logic-of-expectation-potential-information}
The expectation potential information principle imposes
an interesting constraint on information yields from experiment
planning.  Whereas an ``accidental'' observation can produce
a huge $I_p$ yield (e.g. seeing white flowers on a pea plant
yields approximately 68 bits),
the \emph{expected} yield from a planned experiment tends to be
limited to $E(I_p) \to \log{N}$ where $N$ is the
number of distinct models it can distinguish.  Indeed, in the
common situation where we are testing a new model vs. the old
model, every ``good'' experiment becomes at most a 1 bit
hypothesis test.  In this sense, $E(I_p)$ comes very
close to a Boolean logic, in which the scientific method
becomes a series of ``true vs false'' hypothesis tests.
This may seem surprising, given that the metric is based
strictly on observation log-likelihoods, which have no finite limit
(if the model fits the observations poorly, $I_p$ can
grow arbitrarily large).  It is the expectation principle
that strongly constrains the metric: it \emph{can} take into account
the possibility of a hugely surprising experimental outcome,
but by definition assigns it a vanishingly small probability.
Thus the biggest $E(I_p)$ yield comes not from such ``big''
surprises but instead from ``little'' surprises (i.e. with 50\%
probability).

However, it's worth noting that there are some exceptions to
this simplistic ``Boolean'' pattern.  First, it doesn't
apply automatically to \emph{any} arbitrary model proposal, even
if one can design an experiment that would test it.
Unless there is potential information (observations) \emph{favoring}
the new model over the old model, there is no justification
for assigning it a large (50\%, uninformative) probability.
Indeed, arbitrary complex models are likely to be assigned
small prior probabilities (simply to satisfy normalization
of their priors; c.f. Occam's razor), and thus to receive
correspondingly small $E(I_p)$ values.  Second,
the risk of ``technical failure'' problems in the
experiment design can reduce either the total
$E(I_p)$ or the \emph{information rate} far below 1 bit
(e.g. see the Technical Failure example in section 2.3).
Third, if we are computing a \emph{targeted information metric}, it
may be reduced far below 1 bit (even for a good experiment)
if there is strong uncertainty that the observation is
relevant to the target of interest (e.g. if RoboMendel is interested
in inheritance, he initially has uncertainty whether
\emph{Wh} is actually a heritable trait).  On the other hand,
the targeted information yield might be far \emph{larger} than 1 bit,
if the new model gives predictions not only for the one observable
being tested, but for many other relevant observables as well
(for example, RoboMendel's transmission model yields improved
prediction power not only for \emph{Wh x Pu} but for many other traits).
Fourth, one could of course test \emph{many} plausible models
in one experiment design, which would yield an $E(I_p)$
much greater than one bit.  As an extreme example, imagine
a botanist planning a return expedition to a new continent,
based on having discovered many new species on his two previous
expeditions there.  In effect the ``new models'' being tested
are the set of all possible new botanical species, and
the previous expeditions provide reasonable confidence that
the ``experiment design'' can detect many of these ``big surprises''
(i.e. accurately observe their surprising traits).  In
principle, we could empirically estimate a very large
$E(I_p)$ for the return expedition, from the \emph{observed}
$I_p$ yields from the previous expeditions.

\section{Methods Summary}
\label{exptplan:methods-summary}
We have already presented the $E(I_p)$ computations at
length in the theory section (Section 2), and all details of
the calculations are available as source code for the information
metric values presented in this paper (available from
{https://github.com/cjlee112/darwin}).  Here we simply
provide a quick reference summarizing the calculations for
Section 3 so readers can understand them ``at a glance''.

\textbf{Expectation potential information}: $I_p$ is an
empirical (sample-based) estimator of the relative entropy
of the true (but unknown) observation likelihood
$\Omega(X)$ vs. the
model observation likelihood $\Psi(X)$.
To compute its expectation value
for a set of possible likelihood distributions $\omega_i(X)$
with estimated mixture weights $p(\omega_i|\Psi)$, we compute
the potential information yield for each possible experiment
``outcome'' $\omega_i(X)$ vs. the original mixture model
which we define as $\psi_{PL}(X)=\sum_i{p(\omega_i|\Psi)\omega_i(X)}$:
\[
E(I_p) = \sum_i{p(\omega_i|\Psi) D(\omega_i(X)||\psi_{PL}(X))}
\]
where the relative entropy $D(\omega_i(X)||\psi_{PL}(X))$ is
defined as usual
\[
D(\omega_i(X)||\psi_{PL}(X))
= \sum_X{\omega_i(X)\log{\frac{\omega_i(X)}{\psi_{PL}(X)}}}
\]
for a discrete variable $X$ (or replaced by an integral
over $X$ if it is a continuous variable).

\textbf{Targeted information metric}: to restrict the information metric
to a target phenomenon (e.g. genetic inheritance), we re-weight
the $E(I_p)$ for a given observable $X$ by the
probability $\tau_X$ that $X$
is indeed an exemplar of that target
phenomenon.  For example, at the beginning of the RoboMendel
experiments, there is strong uncertainty as to whether \emph{Wh}
is in fact a genetic trait (that can be inherited, as opposed to
an environmental phenomenon).  We calculate the expectation
targeted information value as
\[
E(I_p(X|\Psi,\tau)) = \tau_X E(I_p(X|\Psi))
\]

\bibliography{infoevo}

\end{document}